\documentclass[aps,prc,twocolumn,floats,floatfix,superscriptaddress,showpacs]{revtex4-1}
\usepackage{graphicx,amsmath,amssymb,bm,multirow}
\usepackage{booktabs}

\newcommand{\be}[1]{\begin{equation}\label{#1}}
\newcommand{\ee}{\end{equation}}

\begin{document}

\title{Probabilistic Cellular Automata for Granular Media in Video Games}
\author{Jonathan Devlin}
\email[E-mail:~]{devlinj1@wit.edu}
\affiliation{Wentworth Insititue of Technology, 550 Huntington Ave, Boston, MA 02115, USA}

\author{Micah D. Schuster}
\email[E-mail:~]{schusterm@wit.edu}
\affiliation{Wentworth Insititue of Technology, 550 Huntington Ave, Boston, MA 02115, USA}


\begin{abstract}
    Granular materials are very common in the everyday world.
    Media such as sand, soil, gravel, food stuffs, pharmaceuticals, etc. all have similar irregular flow since they are composed of numerous small solid particles.
    In video games, simulating these materials increases immersion and can be used for various game mechanics.
    Computationally, full scale simulation is not typically feasible except on the most powerful hardware and tends to be reduced in priority to favor other, more integral, gameplay features.
    Here we study the computational and qualitative aspects of side profile flow of sand-like particles using cellular automata (CA).
    Our CA uses a standard square lattice that updates via a custom, modified Margolus neighborhood.
    Each update occurs using a set of probabilistic transitions that can be tuned to simulate friction between particles.
    We focus on the look of the sandpile structure created from an hourglass shape over time using different transition probabilities and the computational impact of such a simulation.
\end{abstract}

\maketitle

\section{Introduction}

There are many computational methods that can be used to compute physical phenomena.
Modeling a complex system first from principles is ideal in many scientific fields, where physical accuracy is paramount.
However, for video games and computer graphics, performance often trumps accuracy. Provided that the result looks realistic enough, a simple physical simulation can be enough to create the illusion of a more complex interaction without the computational cost.
This work looks to simulate physical phenomena in a realistic manner using cellular automata (CA), as done in games such as Noita \cite{noita}, Rimworld \cite{rimworld}, and Oxygen Not Included \cite{ONI}.
Our model uses probabilistic transitions and a modified Margolus neighborhood \cite{toffoli87} to simulate the flow of granular media through an hourglass.

The most common two-dimensional sandpile simulation is the Abelian model \cite{bak87}.
It is simulated from a top-down perspective following sand topping rules based on height.
From a mathematical perspective, this simulation is often used to study concepts such as self-organized criticality, as an examination of how complexity arises in nature \cite{bak95}, rather than for video games or computer graphics.
Side profile sandpile simulations \cite{cervelle07,dennunzio09,goles93}, however, are more relevant for video games.
Noita’s primary game mechanics, for example, use a side profile sandpile model as the basis for their fluid, gas, and solid simulations \cite{purho2019}.
This leads to interesting interactions between the game world and the player, where it is possible to directly manipulate the environment in response to a player's actions.

At a fundamental level, CA are simplified simulations of physical phenomena where time and space have been discretized and physical quantities are represented by a finite set of values.
The problem domain is defined as a collection of cells whose properties are updated over time via a predetermined set of rules. The rules for updating the cell typically consider local information only, i.e. only nearby cells are considered, known as a neighborhood.

The concept was originally developed by Stanislaw Ulam and John von Neumann in the 1940’s at Los Alamos National Laboratory for work on crystal growth and self-replicating robots \cite{vonNeumann66}. Further work was done by scientists and mathematicians throughout the rest of the century.
Stephen Wolfram, in the 1980’s, explored CA in the context of mathematics, physics, biology, and chemistry \cite{wolfram84a,wolfram84b,wolfram85a,wolfram85b,wolfram86a, wolfram86b}.
His book on this research, A New Kind of Science \cite{wolfram02}, is an important starting point for learning about physical modeling using CA.

A common example of CA is Conway’s Game of Life (GOL) \cite{gardner70}. GOL uses a two-dimensional square lattice where each cell can be alive (1) or dead (0). Updating the cells is defined by five rules that represent concepts such as overpopulation or birth.
Although the simulation is simple, it displays surprisingly complex behavior.
The typical structures that emerge from a purely random starting grid consist of groups of static cells, groups that oscillate between two or more states, and groups that traverse the grid.
The GOL model shows that when dealing with CA, simple starting rules can lead to complex behavioral results.

CA modeling of physical systems provides an excellent computational framework for modeling the macroscopic behaviors of various phenomena provided that the update rules can approximate the microscopic properties.
Therefore, CA preserves features such as simultaneous motion, local interactions, and time reversal symmetry for non-probabilistic rules. In addition, the lattice structure of the simulation is ideal for modern parallel computers, where one can perform large simulations with minimal hardware.

Common applications of CA to model complex systems range from fluid dynamics to traffic congestion.
Notable CA such as lattice gas automata for fluid dynamics \cite{hardy76,frisch86,wolf04}, as well as chemical reactions, such as the Belousov-Zhabotinsky reaction \cite{zhabotinskii64} using the Greenberg-Hastings Model \cite{greenberg78}, are examples of CA applied to real world applications.

The flexibility that CA offers in video game environments, while keeping calculation time to a minimum, provides a simple way to represent complex interactions.
In this context, otherwise prohibitively expensive computations, such as interactions between gases, liquids, and solids, can be implemented following simple rules that produce the desired complex behavior.
This allows games to feel more realistic without the need to solve the complex equations that govern these processes.

This work aims to examine the computational and qualitative features of a modification to the standard side-profile sandpile simulation.
We start by describing our sandpile model and how we approached simulating sand particles falling in an hourglass.
Then we will present the results and computational performance of our simulations using probabilistic transitions.
Finally, we conclude with our perspective on future work and modifications to the model.

\section{Sandpile Model}

\begin{figure}[h]
    \begin{center}
        \includegraphics[width=0.2\textwidth]{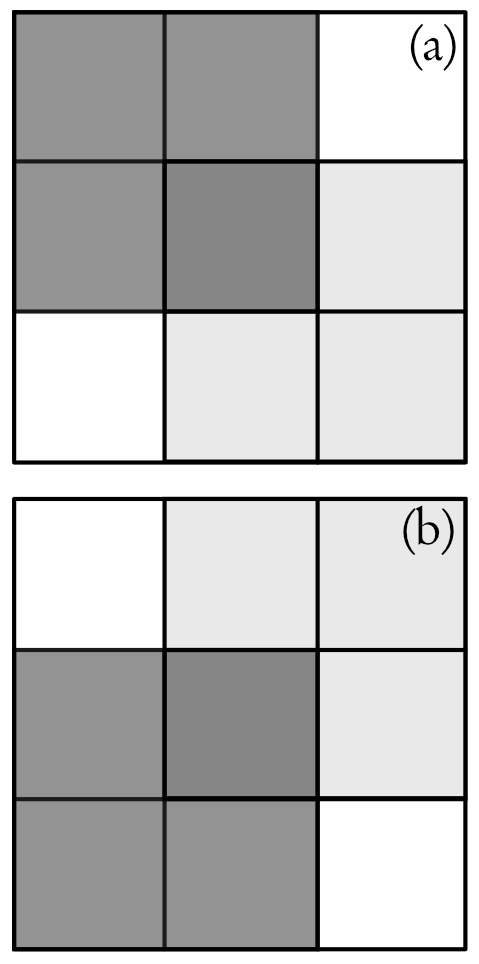}
    \end{center}
    \caption{Standard alternating 2x2 block Margolus neighborhood, panel $(a)$, and our addition to the neighborhood, panel $(b)$. The standard neighborhood alternates between the dark and light 2x2 blocks on successive iterations. With our addition, the simulation alternates between four 2x2 blocks, two in panel $(a)$ and two in panel $(b)$.\label{fig:marg}}
\end{figure}

Our sand pile model uses a modified Margolus neighborhood, shown in Fig. \ref{fig:marg}.
The standard version of this neighborhood is first partitioned into 2x2 blocks of cells on the lattice.
On successive iterations, the 2x2 block is shifted one cell in each dimenion before evaluation.
Although any block CA that uses a Margolus neighborhood can be computed using a larger neighborhood, the amount of states to consider during each iteration increases rapidly. 
Our change to the standard Margolus neighborhood adds an additional shift to the 2x2 block.
This creates a cyclical pattern over four iterations.
The first two are the standard Margolus update, the second two shift the neighborhood as shown in Fig. \ref{fig:marg}b.

As an example of this update process, consider $(x,y)$ to be upper-left corer of the 2x2 block on the first iteration.
The upper-left corner in successive iterations would be:
\begin{itemize}
    \item Iteration 2: $(x+1,y+1)$
    \item Iteration 3: $(x,y+1)$
    \item Iteration 4: $(x+1,y)$
\end{itemize}
Note there may be cells on the domain boundary that are updated on every other iteration.
This is typically not an issue for our simulations since the interior cells, where the dynamics occur, will be updated properly.

The motivation for this change stems from the toppling behavior of the particles when interacting with static boundary cells, whether the finite boundary of the simulation or lattice features for the particles to interact with.
The standard neighborhood causes sand particles to topple in one direction only until they can no longer fall in that specific direction.
When forming a pile, one side builds to the top before the toppling behavior switches to the other slope.
When interacting with static features within on the lattice, this can cause particles to stick to left hand slopes.
Our modified neighborhood forces partiles to alternate topple direction, yielding a more interesting simulation.

When modeling a physical system, it is important to consider the real world properties of the media being simulated.
Accounting, in some way, for the microscopic motion of our particles will lead to a more realistic and insightful result.
As sand pours through an hourglass, the grains contact each other which, due to nonuniform particle shapes, prevents a smooth and consistent flow.

\begin{figure*}
    \begin{center}
        \includegraphics[width=0.9\textwidth]{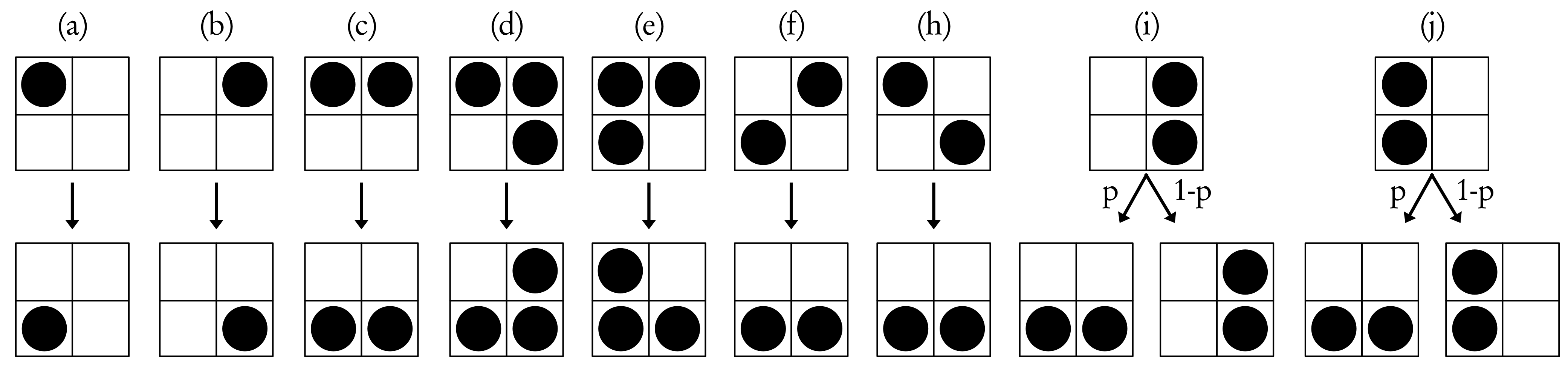}
    \end{center}
    \caption{All transitions that change the particle configuration within the Margolus neighborhood. The two probabilistic transitions (i) and (j) will transition to a new state with probability $p$.\label{fig:trans}}
\end{figure*}

Using a regular square grid, it is impossible to directly simulate nonuniform particles and their interactions.
Instead, we use a set of probabilistic transitions to approximate the effect of friction and other particle interactions.
Fig. \ref{fig:trans} shows all of the transitions that change state.
Most of the transitions, Fig. \ref{fig:trans}(a) - \ref{fig:trans}(h), will always transition in a natural way, falling or toppling down.
The final two transitions, Fig. \ref{fig:trans}(i) and \ref{fig:trans}(j), where the particles are stacked vertically, will only topple with probability $p$.
This represents the effect of irregular shape and friction as the particles move past each other.

These transitions prevent a smooth flow of particles as they begin to stack, which, as shown in the results, ultimately changes the overall piling behavior of the particles as they come to rest.
While we chose two of the transitions to exhibit the probabilistic behavior, for analysis purposes, transitions (c)-(h) could also have associtaed probablities.
This would give seven parameters to adjust to achieve the desired toppling and stacking behavior.
One can even bias the tople direction by choosing different probabilities for the left and right version of the transition.


\section{Results}
We obtained all our results using an hourglass shaped container inside of a 61x61 cell grid.
The edges of the hourglass are fixed to the occupied state and do not change their state over the course of the simulation.
For each iteration, we use the modified Margolus neighborhood, as described above, and the transitions shown in Fig. \ref{fig:trans}.

We start our investigation by examining the deterministic simulation, where the probability, $p$, for all transitions in Fig. \ref{fig:trans}, is set to 1.
Thus, using the same initial condition will always result in the same evolution of the simulation.
We then adjust the probability of the transitions and investigate the differences in the piling behavior.
Computationally, each iteration of the domain took a few milliseconds to calculate, with our visualization dominating the execution time.
For time sensitive applications, such as computer graphics and video games, this style of simulation can easily be translated to production applications.

\subsection{Deterministic Simulation}

\begin{figure*}
    \begin{center}
        \includegraphics[width=0.9\textwidth]{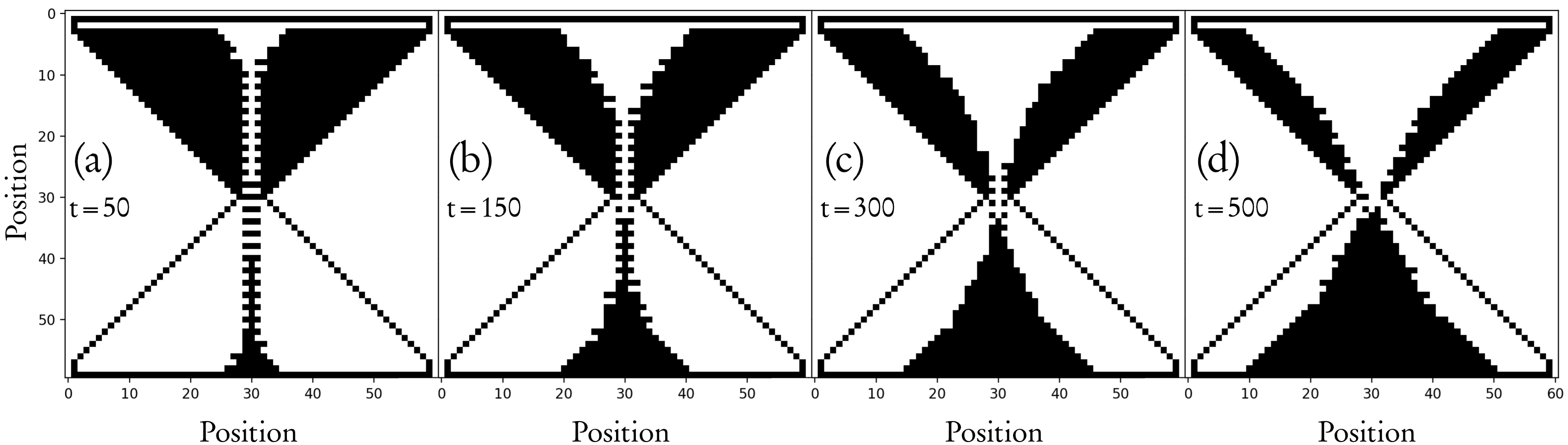}
    \end{center}
    \caption{Deterministic sandpile hourglass simulation time progression from $t=50$ to $t=500$, (a)-(d). When there are no probabilistic transitions the result of the simulation is completely determined by the starting condition. This deterministic simulation acts as the baseline for comparison.   \label{fig:deter}}
\end{figure*}

The deterministic approach fixes every transition with $p = 1$.
Without a probability governing toppling behavior, any given starting condition of sand will result in the same sand position for any given number of generations regardless of how many runs of the simulation are performed.
The deterministic simulation gives us a baseline for which to compare, which is especially can be useful when adding additional interactions between particles.
Fig. \ref{fig:deter} shows the formation of the sandpile over time.

Despite its moderate accuracy representing granular flow, there is a noticeable bias in how the sand fills the hourglass.
The particles alternate filling one slope completely before falling down the other slope.
This pattern alternates as the pile is formed, breaking any attempted immersion as sand falls in an abnormal but constant way.
The unusual way the particles interacted prompted us to add a probabilistic system that reproduced friction-like interactions.

\subsection{Probabilistic Simulation}
For this approach we adjust the probabilities of two transitions, shown in Fig. \ref{fig:trans}(i) and (j).
The piling behavior of the deterministic simulations can then be used as a baseline for comparison when adjusting these probabilities.
Our goal for these simulations is to examine the changes in the structure of the sandpile.

\begin{figure*}
    \begin{center}
        \includegraphics[width=0.9\textwidth]{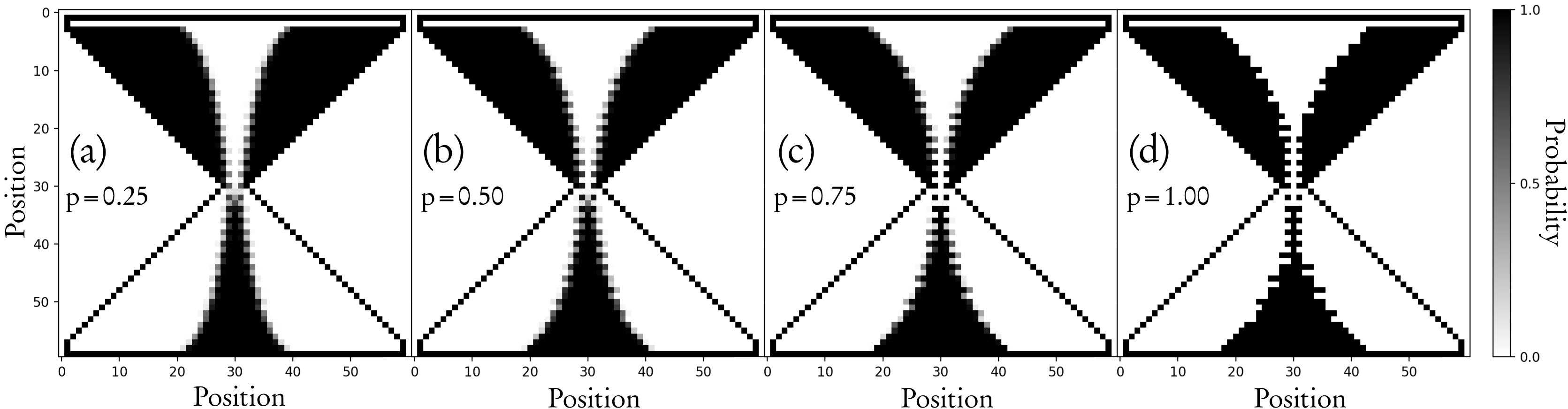}
    \end{center}
    \caption{Sandpile hourglass simulation after 200 iterations using different transition probabilities (see text for details). The colorbar represents the probability that a particle exsists in the associated cell. Lower probability transitions create long, skinny piles compared to higher probabilities. \label{fig:prob}}
\end{figure*}

Fig. \ref{fig:prob} shows how the sand piled on the bottom of the hourglass after 200 iterations in response to four different values of $p$: $0.25$, $0.50$, $0.75$, and $1.00$.
Each simulation was executed 10000 times and the average probability for a particle to exsist in any given cell is shown by the colorbar.
The $p=1.00$ simulation is equivalent to the deterministic simulation and is shown for reference.

The simulations where $p<1.00$ show areas around the pile where sand particles may be present after the 200 iterations.
As the value of $p$ decreases, the sandpile becomes thinner and taller as a result of particles resisting topple.
In effect, this can be interpreted as a friction-like force that keeps the individual grains from sliding past each other.
Over time, the grains will settle as they will eventually come to a stable configuration.

From a qualitative perspective, when $p\leq0.50$, the pile appears more like a viscous fluid rather than a discrete set of grains.
This leads to an upstable looking pile that settles over many iterations.
Values larger than $p=0.50$ tends to give more pleasing granular flow while also providing variation in the piling and toppling behavior.

\subsection{Computational Performance}

To use this simulation in production computer graphics or video games, it must be performant at the desired resolution for the application.
To see the toppling behavior in detail, we used a low-resolution hourglass for our results.
However, in practice, we expect the necessary resolution to be much higher.
Modern video games, for example, typically use resolutions of $1080$p to $4$k to enhance visual fidelity.

By their nature, the speed of any CA simulation will be dependent on the specifics of the computing hardware.
Thus, a direct comparison of the compute times is less relevant than the scaling behavior as we increase the resolution of the simulation.
Since CA are compute heavy, the only relevant piece of hardware is the CPU.
For all of our calculations we used a $4.0$GHz Intel Core i7 6700K.
Table \ref{tab:compute} shows the compute times for five common resolutions, their cell count, and time scaling.
We compute the time for the calculation of one iteration averaged over 100 individual runs.
The scaling uses the 640x480 resolution as a baseline.

\begin{table}[h]
    \centering
    \begin{tabular}{l r r r}
        \hline
        Resolution~~~ & ~~~Time (ms) & ~~~~~~~~~Cells & ~~~Scaling \\
        \hline
        640x480       & 3.08         & 307,200        & 1.00       \\
        800x600       & 4.57         & 480,000        & 1.45       \\
        1920x1080     & 21.51        & 2,073,600      & 6.98       \\
        2560x1440     & 38.84        & 3,686,400      & 12.61      \\
        3840x2160     & 85.88        & 8,294,400      & 27.88      \\
        \hline
    \end{tabular}
    \caption{Compute times, number of cells, and time scaling for different simulation resolutions.
        All times represent computing one iteration of the simulation and are the average of 100 individual runs.
        640x480 is the baseline used to compute scaling as the resolution increases.}
    \label{tab:compute}
\end{table}

The time scaling follows the increase in cells nearly identically.
This is because the time per iteration is solely dependent on performing the Margolus neighborhood calculations, which itself depends on the total number of cells in the domain.
The memory footprint is negligible for this style of simulation because we can use a single byte to represent the occupation of each cell.

For a typical video game, where a simulation like this may take place, one has approximately 16 milliseconds for all calculations in one frame.
Thus, when using a smaller grid, more time can be allocated to other computations, such as physics, input, animation, etc.
Typically, these types of physical simulations do not associate one cell with one screen pixel.
Rather, lower resolutions are used to approximate liquids, gases, and granular media that are relevant to the specific game.
For example, Terraria \cite{terraria} uses 16x16 pixel tiles to represent the game world.
When computing its CA fluid simulations, this greatly reduces the number of cells to calculate on each frame.
Overall, the highest resolution possible for the given context is desirable.

\section{Conclusions}

We have shown the characteristics of sandpiles using a probabilistic CA model.
The toppling behavior and final structure are largely dependent on the chosen probabilities and can be adjusted to create a friction-like interaction between particles.
This causes elongated piles and longer settling times as the probability to topple decreases.
Using our modified Margolus neighborhood, we get a complex and interesting flow of particles that reflect real world motion.
Computationally, our method provides a balance between fast computation and realistic flow, even at high resolution.

In game development, where framerate is paramount, using CA for physical simulations allows for fast calculations while maintaining a high framerate.
In addition, simulating multiple types of materials and their interactions adds very little computational overhead while generating complex behaviors.
The system we modeled could handle up to 4k resolution while maintaining a reasonable execution time per iteration. Most games, however, do not simulate each pixel.
Instead, small regions of pixels represent one cell of data. Thus, a 4k screen region can be simulated with less cells than our 640x480 calculation.

When implementing new rules for new materials, there are possibilities for refining the calculations.
Breaking up the domain into large groups of cells and preventing updates on groups that are unlikely to change value, such as at the domain boundary, can significantly speed up the calculations.
For our simulations, the domain was small enough that this type of optimization was not necessary.
However, for high resolution production applications, this represents an efficient way to reduce the calculation time per iteration.
For the future, we intend to add additional materials that interact probabilistically using our modified neighborhood.
This leads to increased implementation complexity while, at the same time, gives our simulation more interesting behaviors.
In addition, we intend to design a fully configurable toy simulation that can be used and expanded upon in modern game engines.

\section*{Acknowledgments}
This work was performed with the support of the Department of Computer Science and Networking at Wentworth of Technology.

%

\end{document}